\documentclass[twocolumn,prl,citeautoscript,superscriptaddress,8pt]{revtex4-2}

\usepackage{graphicx}
\usepackage{gensymb}
\usepackage{color}
\usepackage[dvipsnames]{xcolor}
\usepackage{float}
\usepackage{upgreek}
\usepackage{bm}
\usepackage{amsmath,amssymb}
\usepackage{array}
\usepackage{hyperref}
\newcolumntype{M}[1]{>{\centering\arraybackslash}m{#1}}
\newcommand{\ket}[1]{\ensuremath{\left| #1 \right\rangle}}

\hypersetup{urlcolor=blue, colorlinks=true, linkcolor=blue,citecolor=blue, linktocpage=true}

\begin{document}

\title{Radiofrequency receiver based on isotropic solid-state spins}

\author{Islay O. Robertson}
\affiliation{School of Science, RMIT University, Melbourne, VIC 3001, Australia}

\author{Brett C. Johnson}
\affiliation{School of Science, RMIT University, Melbourne, VIC 3001, Australia}

\author{Giannis Thalassinos}
\affiliation{School of Science, RMIT University, Melbourne, VIC 3001, Australia}

\author{Sam C. Scholten}
\affiliation{School of Science, RMIT University, Melbourne, VIC 3001, Australia}

\author{Kevin J. Rietwyk}
\affiliation{School of Science, RMIT University, Melbourne, VIC 3001, Australia}

\author{Brant Gibson}
\affiliation{School of Science, RMIT University, Melbourne, VIC 3001, Australia}

\author{Jean-Philippe Tetienne}
\email{jean-philippe.tetienne@rmit.edu.au}
\affiliation{School of Science, RMIT University, Melbourne, VIC 3001, Australia}

\author{David A. Broadway}
\email{david.broadway@rmit.edu.au}
\affiliation{School of Science, RMIT University, Melbourne, VIC 3001, Australia}

\begin{abstract} 
Optically addressable solid-state spins have been proposed as robust radiofrequency (RF)-optical transducers sensitive to a specific RF frequency tuned by an external static magnetic field, but often require precise field alignment with the system's symmetry axis. 
Here we introduce an isotropic solid-state spin system, namely weakly coupled spin pairs in hexagonal boron nitride (hBN), which acts as an RF-optical transducer independent of the direction of the tuning magnetic field, allowing greatly simplified experimental design.
Using this platform, we first demonstrate a single-frequency RF receiver with frequency tunability from $0.1$ to $19$\,GHz.
We next demonstrate an instantaneous wideband RF spectrum analyser by applying a magnetic field gradient to encode RF frequency into spatial position.
Finally, we utilise the spectrum analyser to detect free-space-transmitted RF signals matching the strength and frequency of typical Wi-Fi signals.   
This work exemplifies the unique capabilities of isotropic spins in hBN to operate as RF sensors, while circumventing the challenging requirement of precisely aligned magnetic fields facing conventional solid-state spins.
\end{abstract}

\maketitle


Radiofrequency (RF) signal detection and analysis techniques play important roles in various fields, including wireless communication, medicine, and navigation.
Traditional electronics-based approaches involve digitizing the RF signal (directly or after frequency down-conversion) with an analog-to-digital converter, which is then analysed via digital signal processing to extract the relevant information. 
However, this approach has operational limitations in environments with greatly varying signal strengths, especially in the presence of high-power RF signals which can lead to saturation and even damage. 
Quantum sensors have the potential to replace electronics-based RF receivers in extreme scenarios, as they are often highly sensitive and not subject to damage under high-power signals\,\cite{DegenRMP2017}.
An example quantum sensor for RF signal detection is Rydberg atoms, which have achieved high sensitivities over broad spectral ranges using both passive and active (frequency mixing) methods\,\cite{CoxPRL2018, AndersonIEEEAESMag2020, MeyerPRApp2021}.
While significant advancements have been made towards the operation and miniaturization of these sensors, there are challenges with engineering efficient RF-atom coupling and complex optical setups with multiple tunable lasers for resonant optical driving.
Solid-state spin defects with optically addressable magnetic resonances are a competing platform for RF sensing, and for many applications, are favoured for their robust, simple, and compact operation\,\cite{ChipauxAPL2015, WangNatComms2015, StarkNatComms2017, MagalettiCommEng2022, WangPRX2022, WangSciAdv2022, OgawaAPL2023, WangIEEEMWTheoryTech2024, ZhangOptExp2024}.
Similar to Rydberg atoms, both active and passive RF detection modalities have been explored with solid-state spin defects in a variety of different materials\,\cite{WangNatComms2015, JiangSA2023, RizzatoNatComms2023, PatricksonNPJQI2024}.

Active quantum RF sensors typically employ homo- or heterodyne techniques to increase sensitivity\,\cite{MeyerPRApp2021, MeinelNatComms2021}, however, these methods bring additional technical complications requiring extra reference RF signals.
On the other hand, passive detection, which will be the focus of this work, requires no additional input as the quantum sensor directly interacts with the input RF signal.
In this modality, the spins act as RF-optical transducers by encoding the amplitude of the RF input into photoluminescence (PL) intensity which is detected and digitised via optical detectors and electronics, thus providing electric isolation between the RF input and the readout electronics\,\cite{WangNatComms2015, AppelNJP2015, ShaoPRApp2016, HorsleyPRApp2018, BaiAPL2022, CaiAPL2024}.
The PL intensity is modulated through spin-dependent photodynamics by resonant driving of spin transitions from a matching RF signal.
This mechanism enables frequency-selective detection of RF signals, where the resonance frequency can be tuned by an applied static magnetic field via the Zeeman effect.
However, most common solid-state spin defects have spin transitions in $S \geq 1$ manifolds with a spin quantization axis set by the symmetry axis of the defect. 
This intrinsic directionality means the applied magnetic field must be well aligned with the symmetry axis in order to preserve the optical contrast\,\cite{KoehlNat2011, TetienneNJP2012, MuPRL2022}, which poses significant challenges when designing frequency-tuneable RF receivers.

In this paper, we consider the use of weakly coupled spin pairs, which have spin-$1/2$-like behaviour with an isotropic Zeeman response and an optical response independent of the direction of the applied magnetic field\,\cite{BoehmePRB2003}, a property which relaxes some restrictions for sensing applications\,\cite{GengNatComms2023}.
Namely, we employ recently reported spin pairs in the van der Waals material hexagonal boron nitride (hBN) \cite{ChejanovskyNatMat2021, SternNatComms2022, GuoNatComms2023, ScholtenNatComms2024}, and show this system can act as an RF-optical transducer.
To this end, we realise two different systems: (1) a simple single-frequency receiver in a uniform field for the measurement of amplitude-modulated RF signals, and (2) a spectrum analyser for instantaneous multi-frequency analysis using a magnetic field gradient to encode RF frequency into spatial position. 
We demonstrate measurements of directed RF signals with both cable and free-space transmission. 
This work establishes the hBN spin pairs as a promising system for RF signal detection and analysis, and furthermore, prompts exploration of other magnetic field gradient based applications \cite{AraiNatNano2015, AmawiNPJ2024, GuoNPJ2024}.


\begin{figure}[t]
    \centering
    \includegraphics{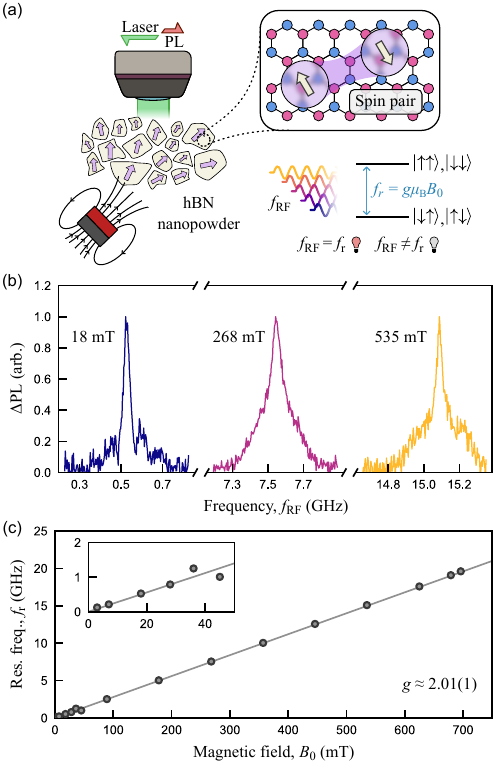}
    \caption{
    \textbf{Frequency tuneability of spin pairs in hBN.}
    (a)~Schematic representation of the experiment. 
    hBN nanopowder with ensembles of spin pairs is illuminated with a $532$\,nm laser.
    The spins self-align to arbitrarily applied magnetic fields and are subject to target radio-frequency (RF) signals of frequency $f_{\rm RF}$. 
    When $f_{\rm RF}$ matches the resonance frequency $f_r$ of the effective two-level system (pure triplet $\ket{\uparrow\uparrow}$ and $\ket{\downarrow\downarrow}$, and singlet-triplet mixtures of $\ket{\uparrow\downarrow}$ and $\ket{\downarrow\uparrow}$), there is a change is photoluminesence (PL) intensity.
    (b)~Example ODMR spectra for the spin pair ensemble for different values of the applied magnetic field ($B_0 = 18, 268, 535$\,mT).
    (c)~Plot of measured central ODMR resonance frequency ($f_r$) as a function of applied magnetic field (grey markers). 
    The solid grey line is a linear fit with the relationship $h f_r = g \mu_B B_0$ giving $g=2.01(1)$.
    Inset, a zoom in on the first few points.
    The first point is at $B_0\approx 3$\,mT ($f_r\approx100$\,MHz).
    }
    \label{fig1}
\end{figure}

\begin{figure*}[tb]
    \centering
    \includegraphics{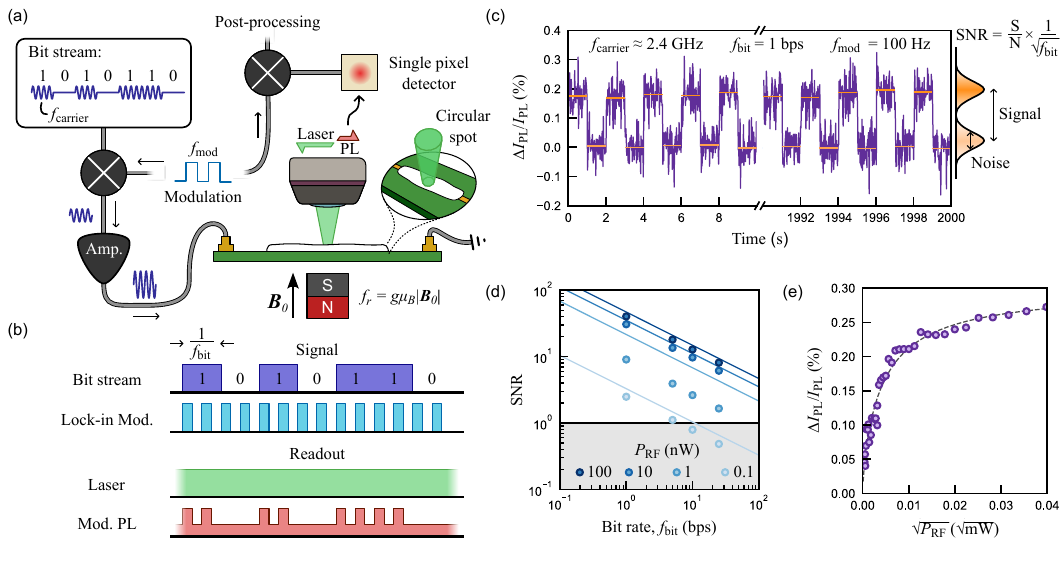}
    \caption{
    \textbf{Single-frequency RF signal receiver.}
    (a)~Schematic representation of the setup. 
    A lock-in modulated ($f_{\rm mod}$) bit stream ($f_{\rm bit}$) carried by the RF wave ($f_{\rm carrier}$) is amplified and directed to the PCB. 
    The PL collected in epifluorescence is directed to a single pixel detector and the output demodulated at $f_{\rm mod}$ to reveal the data stream. 
    (b)~Measurement sequence for extracting the data bit stream. 
    The laser is continuously applied throughout the measurement. 
    Lock-in modulation generates a reference for common-mode rejection.
    (c)~Demodulated PL time trace with a periodic $f_{\rm bit}=1$\,bps bit stream input signal with a $f_{\rm carrier} = 2.4$\,GHz carrier frequency. 
    SNR is calculated as a function of the bit rate by individually averaging the on and off pulses.
    Lock-in modulation frequency for the measurement was $f_{\rm mod} = 100$\,Hz.
    (d)~SNR for different peak RF powers plotted as a function of bit rate. 
    Solid lines are the theoretical expected SNR values in the photon shot noise limit.
    (e)~ODMR contrast plot versus the peak RF power.
    }
    \label{fig2}
\end{figure*}

In our experiments, ensembles of spin pairs in hBN nanopowder are optically illuminated by a $532$\,nm laser and the resulting spin-dependent PL collected [Fig.~\ref{fig1}(a)].
The spin-spin interactions (dipole-dipole and exchange) of the spin pairs are insufficient to cause significant splitting between resulting singlet and triplet states. 
An effective two-level system represents the simplified energy level structure of the spin pair where the two states have varying degrees of singlet character leading to different PL intensities\,\cite{RobertsonArxiv2024}.
The separation of the two states can be tuned by applying a uniform magnetic field $B_0$, giving a splitting of $h f_r = g \mu_B B_0$ where $h$ is Planck's constant, $\mu_B$ is the Bohr magneton and $g$ is the $g$-factor.

Transitions between the two states of the spin pair can be measured via optically detected magnetic resonance (ODMR) by sweeping the frequency of an applied RF field, a resonant transition (at frequency $f_r$) corresponding to an increase in PL giving positive ODMR contrast. 
We have observed ODMR contrast over a large frequency range of $f_r = 0.1-19.5$\,GHz ($B_0 = 3-700$\,mT), three representative examples being shown in Fig.~\ref{fig1}(b).
In our measurements the contrast at saturation is mostly constant throughout this frequency range, varying between $0.1$ and $0.5$\,\%.
The spin$-1/2$ nature of the spin pair is preserved over this field range as indicated by the linear relation to the applied magnetic field with $g \approx 2.01(1)$ [Fig.~\ref{fig1}(c)].


Having demonstrated the easy frequency tunability of the spin pairs in hBN, we now explore their ability to function as a single-frequency RF receiver. 
Here, the goal is to receive data encoded into the amplitude of an RF field at a single known RF carrier frequency $f_{\rm carrier}$.  
This configuration is achieved by affixing a permanent magnet to a printed circuit board (PCB) that contains an RF waveguide on the opposing side with hBN nanopowder drop cast directly onto the waveguide [Fig.~\ref{fig2}(a)].
The laser is focused onto the powder and the PL is collected and filtered before collection on a detector.
The field $B_0$ experienced by the spins can be considered uniform over the interrogated volume, with a value set by the magnet-PCB distance, here $B_0 \approx 86$\,mT. 
The RF carrier frequency is chosen to meet the resonance condition, $f_{\rm carrier}=f_r\approx 2.4$\,GHz.
The test signal is a data bit stream (0 and 1) encoded into the RF amplitude at a clock frequency $f_{\rm bit}$. 
A second amplitude modulation at frequency $f_{\rm mod}$ allows us to perform lock-in demodulation in the optical signal to remove common-mode fluctuations [Fig.~\ref{fig2}(b)]. 
Note, in future implementations this lock-in modulation could be achieved by modulating $B_0$ with an electromagnet instead.   
The doubly-modulated RF signal (peak power $P_{\rm RF}$) is amplified (50\,dB gain) and directed to the waveguide, which acts as the RF-to-spin interface.

We first demonstrate the hBN quantum receiver with a test data bit stream simply alternating between $0$ and $1$ each clock cycle with $f_{\rm bit} = 1$\,bps, which is measured over a $2000$\,s duration [Fig.~\ref{fig2}(c)].
The $0.2$\% average on/off (1/0) signal contrast is commensurate with the ODMR contrast, and is maintained throughout the entire measurement indicating the detection scheme is robust.
We quantify the clarity of signal reception by determining the signal-to-noise ratio (SNR) as a function of $f_{\rm bit}$ by considering the distribution of PL values [Fig.~\ref{fig2}(c), right].
Increasing $f_{\rm bit}$ from $1$ to $25$\,bps reduces the SNR from $40$ to $8$, matching well with expected relationship from the photon shot noise limit [Fig.~\ref{fig2}(d), solid lines]. 
This relationship also scales as a function of the input RF power. 
Indeed, the ODMR contrast ${\cal C}$ increases as ${\cal C}\propto\sqrt{P_{\rm RF}}$ until saturation which occurs at $P_{\rm sat} \approx 4$\,mW [Fig.~\ref{fig2}(e)]. 
As a consequence, the RF power limits the maximum resolvable bit rate (defined as $f_{\rm bit}$ at $\text{SNR} = 1$) which increases from $10$\,bps at $P_{\rm RF} = 0.1$\,nW to $2$\,kbps for $P_{\rm RF} = 100\,$nW (extrapolated from the expected behaviour).  
We define the sensitivity from the RF power with $\text{SNR} = 1$ for a $1$\,bps signal, which gives $\eta_{\rm RF} = 70\,$pW$/\sqrt{\rm Hz}$.
While this experiment validates the concept of using spin pairs in hBN as an RF signal receiver and optical transducer, the low sensitivity (and hence low bit rate) are far below current technology.
Later we will discuss how there is ample opportunity for improvement.


\begin{figure*}[t]
    \centering
    \includegraphics{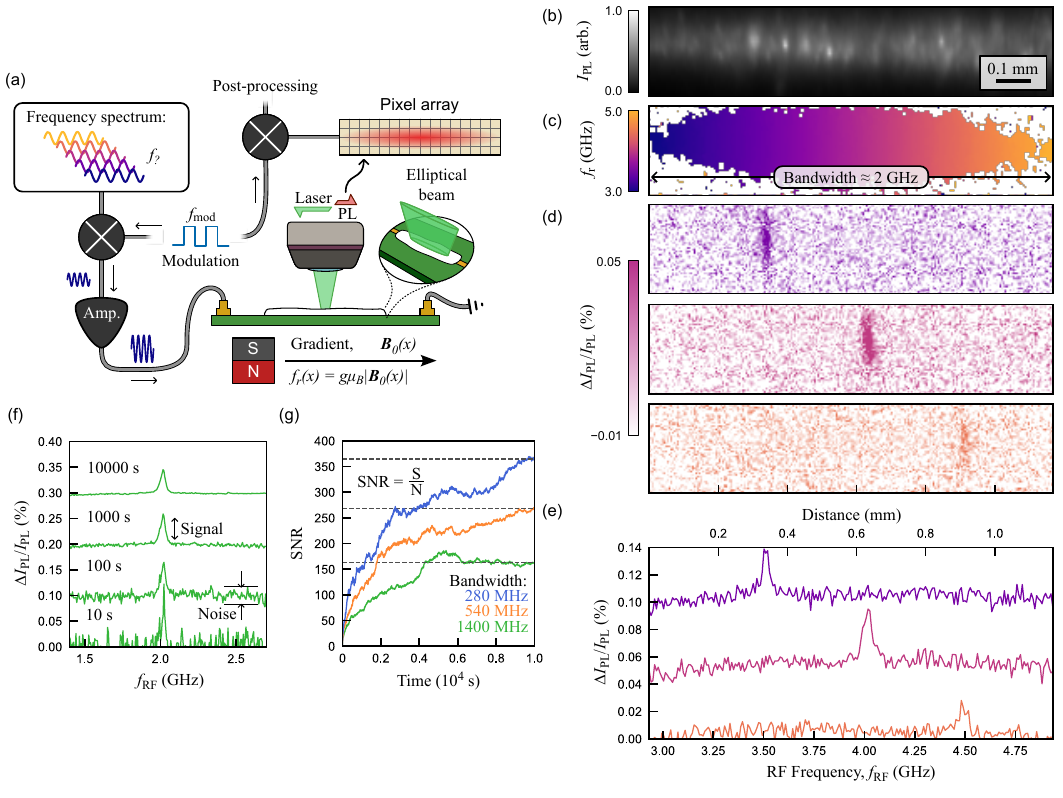}
    \caption{
    \textbf{Instantaneous RF spectrum analyser.}
    (a)~Schematic representation of the setup. 
    RF signals with an unknown spectral content are lock-in modulated ($f_{\rm mod} = 50$\,Hz), amplified and directed to the PCB. 
    A magnetic field gradient assigns spatial positions to the resonant frequencies of the spin pairs.
    Elliptical illumination (green) and epifluorescent collection (red) by an objective is directed to a pixel array for imaging.
    (b)~PL image of the hBN nanopowder. 
    (c)~Corresponding map of the spin resonance frequency under a magnetic field gradient, obtained by ODMR spectroscopy.
    (d)~PL change maps for three different input RF frequencies ($1$\,mW) at $f_{\rm RF}=3.5, 4.0, 4.5$\,GHz.
    (e)~Spectra of the RF input derived from the spatial maps in (d) using the frequency map in (c). 
    (f)~Spectra of the RF input (single frequency $f_{\rm RF}=2.0$\,GHz) integrated for increasing periods of time.
    (g)~SNR as a function of time for three different laser post sizes, corresponding to bandwidths of $1400$\,MHz (green), $540$\,MHz (orange), and $280$\,MHz (blue).
    }
    \label{fig3}
\end{figure*}

In the previous experiment, the spin pairs were sensitive to a single RF frequency set by the uniform magnetic field $B_0$. 
We now modify the system to enable simultaneous detection of multiple RF frequencies.
This is achieved by applying a magnetic field gradient $B_0(x)$ to produce a spatially varying spin resonance frequency $f_r(x)$, then measured by resolving the PL onto a camera (multi-pixel array)\,\cite{ChipauxAPL2015,MagalettiCommEng2022}. 
To maximise the frequency range, we reshape the laser to have a highly eccentric elliptical cross-sectional area with the major axis along the direction of the magnetic field gradient [Fig.~\ref{fig3}(a)].

A raw PL image of the hBN powder is shown in Fig.~\ref{fig3}(b), the bright region corresponding to the elliptical laser spot.
To determine the resonance frequency of the spin pairs along the laser spot, we perform ODMR spectroscopy and fit the data to determine $f_r$ at each pixel\,\cite{Scholten2021}.
In the resulting $f_r$ map [Fig.~\ref{fig3}(c)], the gradient along the principle axis of the laser spot becomes evident, with $f_r$ varying from $3.0$ to $5.0$\,GHz.
The image bandwidth ($\approx 2$\,GHz) is determined by the laser spot size and imaging field of view ($\approx 1.2$\,mm) and its position along the field gradient determines the frequency range.
The $f_r$ map calibrates the given field of view, relating pixel position to the RF frequency the pixel is sensitive to.
To demonstrate spectrum analysis, we consider three different test RF inputs, each containing a single RF frequency $f_{\rm RF}$,  which can be resolved as areas of enhanced PL in normalised PL images [Fig.~\ref{fig3}(d)]. 
By averaging over the vertical axis of these images and ascribing frequency values to the horizontal axis using the previous $f_r$ map, we convert the PL images into instantaneous spectra of the RF input [Fig.~\ref{fig3}(e)].

To evaluate the performance of the spectrum analyser, we integrate the spectrum of a test fixed-frequency signal for increasing periods of time [Fig.~\ref{fig3}(f)].
The PL contrast ($\approx 0.05\%$) remains constant for the $10$-$10^4$\,s integration times considered, though the noise decreases towards a steady-state due to system instabilities on this timescale as the PL signal is averaged.
Plotting the SNR as a function of the integration time shows the SNR saturates to $160$ after $\approx 6 \times 10^3$\,s for the largest field of view (bandwidth $1.4$\,GHz) [Fig.~\ref{fig3}(g)].
In this regime, for a $1$\,mW  (0\,dBm) input power, the SNR reaches unity in roughly half a second of integration, which implies in these conditions our spectrum analyser has a sensitivity of $\eta_{1.4\,\text{GHz}} \approx 1$\,mW$/\sqrt{\rm Hz}$.
SNR can be improved by decreasing the field of view (increasing laser intensity) at the cost of lowering the frequency bandwidth. 
For example, by reducing the bandwidth to $\approx 280$\,MHz, an SNR of $160$ is reached in only $1 \times 10^3$\,s. 
Trivially, the per pixel frequency resolution is improved which can be used to better resolve finely separated channels in a given frequency band, though in these experiments the resolution was limited by the linewidth of the spin resonance ($\approx30$\,MHz).


\begin{figure}
    \centering
    \includegraphics{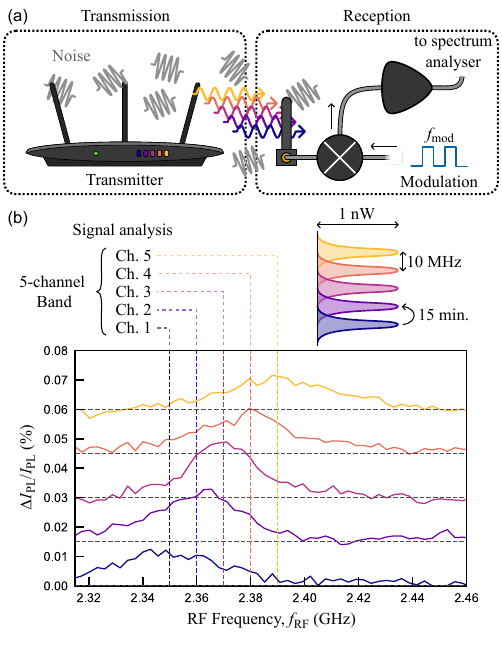}
    \caption{\textbf{Detection of free-space transmission.}
    (a)~Schematic representation of the free-space transmission experiment.
    The setup is separated into transmission and reception arms about one metre apart. 
    The received signal is lock-in modulated ($f_{\rm mod} = 50$\,Hz) and amplified before being directed to the hBN-based spectrum analyser.
    (b)~Measured spectra of artificial signals produced by the transmitter, representative of the frequencies and signal strengths of typical Wi-Fi signals. The power collected by the receiving antenna is $1$\,nW (-60 dBm) and the 5 frequency channels are spaced by $10$\,MHz. Each spectrum was integrated for $15$\,min.
    }
    \label{fig4}
\end{figure}

Finally, we demonstrate the application of the above spectrum analyser for real-world frequency monitoring by recording a signal transmitted through free-space.
A signal generator connected to an antenna forms the transmission arm and directs the signal to a reception antenna connected to the hBN-based spectrum analyser [Fig.~\ref{fig4}(a)].
Currently, our system does not have the sensitivity to detect most data transfer signals, as they are often transmitted over short periods of time with long intervals in between with no signal. 
To compensate for this, we produce an artificial continuous data transfer with the same signal strength as data transfer pulses from a Wi-Fi type-N router (see calibration procedure in SI).  
These test signals were produced for a $5$-channel band with a $10$\,MHz frequency spacing and were transmitted over $\approx 1$\,m. 
The signals were then collected using a standard Wi-Fi broadband antenna and then amplified before being directed to the PCB. 
Each channel was spectrally resolvable within the measurement timeframe of approximately $15$ minutes [Fig.~\ref{fig4}(b)], well above the background noise present in the room. 


In conclusion, the key defining property we have shown which distinguishes the hBN spin pairs are their ease of use resulting from their isotropic response.
This greatly facilitates their frequency tuneability and the application of magnetic field gradients, since no care needs to be taken to maintain a constant magnetic field direction unlike for anisotropic spin systems\,\cite{MagalettiCommEng2022}. 
In particular, the instantaneous bandwidth of the presented spectrum analyser can be easily increased to cover e.g. a $1$-$30$\,GHz range, by simply attaching a smaller magnet close to the interrogated region.

In the presented experiments, the sensitivity in one second of integration was $30$\,nW ($-45$\,dBm) for the single-frequency receiver, and 1\,$\mu W$ ($-30$\,dBm) for the spectrum analyser with $1.4$\,GHz instantaneous bandwidth. 
While this is far below electronics-based technology, hBN spin pairs are a relatively newly discovered system and so we anticipate significant improvements can be gained in the near future. 
For instance, ensemble spin contrast is commonly observed to be quite low ($C < 0.5\%$), but single defects have been observed to have a contrast as high as $C = 60\%$\,\cite{RobertsonArxiv2024}.
An ensemble engineered with this contrast level may be possible, which would improve the sensitivity by two orders of magnitude assuming the sensitivity is limited by photon shot noise.
Likewise, materials engineering to maximise the density of optically active spin pairs, and photonics engineering to maximise photon collection efficiency, should enable significant sensitivity enhancements \,\cite{BarryRMP2020}.  
The sensitivity in terms of RF power could also be improved by optimisation of the signal delivery to maximise the RF magnetic field strength experienced by the spins, for instance using RF cavities.

Finally, the spectral selectivity (receiver modality) and resolution (spectrum analyser) are limited by the natural linewidth of the ODMR peak, which was about $30$\,MHz in our hBN sample. 
Isotopic purification of the hBN host has been shown to improve the ODMR linewidth of boron vacancy defects and could potentially yield similar improvements for the spin pairs\,\cite{HaykalNatComms2022, CluaProvostPRL2023, GongNatComms2024}.
Beyond the passive detection scheme investigated in this paper, it will also be interesting to investigate the effectiveness of active schemes (using additional RF driving signals\,\cite{RizzatoNatComms2023, PatricksonNPJQI2024}) on the spin pair system, which may offer pathways to improve both sensitivity and spectral resolution.

\textbf{Acknowledgments}
This work was supported by the Australian Research Council (ARC) through grants FT200100073, LP210300230 (in partnership with Diamond Defence Pty Ltd), DP220100178, and DE230100192.
I.O.R acknowledges the support of an Australian Government Research Training Program Scholarship.

\bibliography{bib.bib}

\appendix

\section{Experimental apparatus}

Measurements were performed on a custom-built wide-field fluorescence microscope described elsewhere~\cite{ScholtenNatComms2024}. 
In brief, optical excitation from a continuous-wave $\lambda = 532$\,nm laser was focused using a widefield lens to a turret of three possible objectives (4x, 10x, 20x) allowing us to rapidly change the field of view. 
The PL from the sample was separated from the excitation light with a dichroic mirror, filtered using longpass and shortpass filters, and was collected by a scientific CMOS camera. 
The total laser power (going into the objective) was $500$\,mW.

For ODMR spectroscopy and to generate the test signals, RF excitation was provided by a signal generator (Windfreak SynthNV PRO). 
The amplitude modulation (for lock-in modulation and data encoding) was achieved using the list mode of the signal generator and an external trigger. 
The modulated RF was amplified (Mini-Circuits ZHL-16W-43) and connected to the PCB comprising a coplanar waveguide (width of the central strip $100\,\mu$m), terminated by a $50$\,$\Omega$ termination. 
A pulse pattern generator (SpinCore PulseBlasterESR-PRO 500\,MHz) was used to trigger the signal generator and trigger the camera acquisition, enabling lock-in demodulation which was simply achieved by subtracting the frame with lock-in gate off to the frame with lock-in gate on. 

The external magnetic field was applied using a permanent magnet, and the measurements were performed at room temperature in ambient atmosphere. 
The only exception is Fig.~\ref{fig1}(b,c), for which the sample was placed in a closed-cycle cryostat with a base temperature of $5$\,K \cite{LillieNanoLett2020} which allowed us to apply a calibrated magnetic field using the enclosed superconducting vector magnet. 
There was no significant differences in terms of PL and ODMR properties between the room temperature and $5$\,K measurements.
For Fig.~\ref{fig1}(b,c), RF input below $0.7$\,GHz was supplied by a Rohde \& Schwarz SMB100A signal generator with no additional amplifier.
Between $0.7$\,GHz and $6$\,GHz the signal was amplified by a Mini-Circuits HPA-50W-63+.
Above $6$\,GHz, RF signal was supplied by an Agilent MXG N5183A signal generator and amplified by a Mini-Circuits ZVE-3W-183 amplifier.

\section{g-factor determination}

\begin{figure*}
    \centering
    \includegraphics{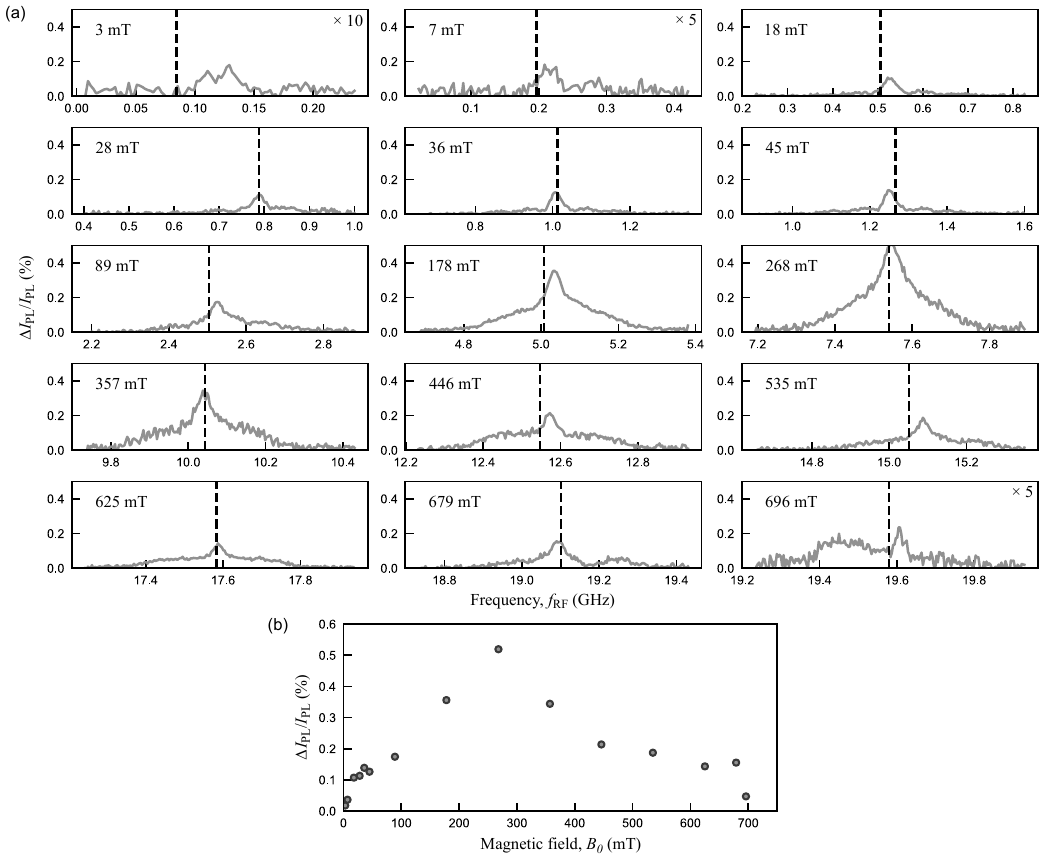}
    \caption{
    \textbf{Additional ODMR data.}
    (a)~Complete ODMR data set for the spin pairs under a range of magnetic fields ($3$-$700$\,mT).
    Dashed lines indicate the expected resonant frequency for a free electron ($g = 2$).
    (b)~ODMR contrast plotted as a function of the applied magnetic field $B_0$.
    }
    \label{si_odmr}
\end{figure*}

The magnetic field dependence was determined by measuring the hBN transition frequency relative to a known magnetic field in a closed-cycle cryostat described above. 
ODMR spectra were taken over the magnetic field range of $3$-$700$\,mT, and this dependence was fit to calculate $g\approx2.01$ in Fig.~\ref{fig1}(c).
The individual spectra from this measurement series are shown in Fig.~\ref{si_odmr}(a).
The average contrast is $0.2$\,\% and varies from $0.02$-$0.5$\,\% [Fig.~\ref{si_odmr}(b)].
This variation in contrast originates from several sources. 
First, at low fields ($\lesssim100$\,mT) mixing between the states of the spin pairs leads to a reduction in contrast\,\cite{TetienneNJP2012} and has been observed consistently across hBN samples\,\cite{SternNatComms2022}.
Second, various signal generators and amplifiers were used for different frequency ranges, modifying the applied RF power across the measurement series. 
Finally, at high fields ($\gtrsim500$\,mT), we experience significant RF transmission loss for the corresponding frequencies resulting in an inherent loss of power and subsequently ODMR contrast.

The line shape of the ODMR spectra is well-described by a two-part function comprised of a Lorentzian and a Gaussian component, consistent with other spin pair systems\,\cite{GengNatComms2023}.
The widths of the two components are on average $30$\,MHz and $300$\,MHz across the range and do not vary significantly otherwise. 

\section{Time-resolved measurements}

\begin{figure}[t!]
    \centering
    \includegraphics{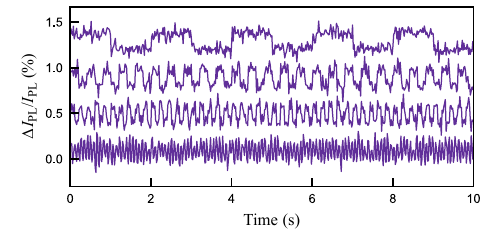}
    \caption{
    \textbf{Additional time-resolved measurements.}
    Time trace of bit streams with bit rates $f_{\rm bit} = 1, 5, 10, 25$\,bps.
    }
    \label{si_sig}
\end{figure}

To supplement the example in main text Fig.~\ref{fig2}(c), here we show [Fig.~\ref{si_sig}] $100$\,nW signals for all explored bit rates ($f_{\rm bit} = 1, 5, 10, 25$\,bps). 
The on/off signal contrast ($0.2$\%) remains constant for all bit rates at this power.
As noted in the main text the maximum resolvable bit rate for the same measurement conditions, extrapolating to ${\rm SNR}=1$, is $2$\,kbps. 
In our current implementation of this experiment, reaching this bit rate is impossible due to thesCMOS camera framerate.
For single-frequency measurements, the camera could be replaced with a photodiode or APD, which lifts this limitation to over $10$\,MHz.
Faster measurements can also be achieved for the spectrum analyzer by using high framerate or lock-in cameras such as the Heliotic helicam C4, which has a framerate of 250 kHz.

\section{Gradient field measurements}

\begin{figure}[t!]
    \centering
    \includegraphics{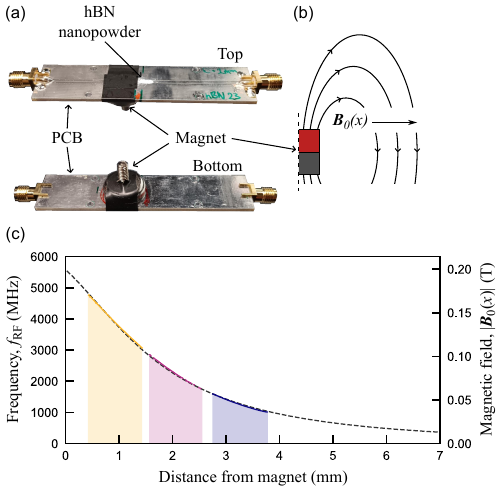}
    \caption{
    \textbf{Calibration of gradient field device.}
     (a)~Gradient field device.
     hBN nanopowder is deposited onto an RF waveguide embedded in a printed circuit board (PCB).
     A permanent magnet is taped to the PCB on one side of the waveguide to supply the magnetic field gradient.
     (b)~Schematic representation of the magnetic field gradient which extends laterally from the magnet.
     (c)~Resonant frequencies measured over the extent of the gradient field device. 
     Measurements are collected in three fields of view (blue, pink, yellow). 
     Simulated magnetic field gradient to match the measured data (black).
    }
    \label{si_gradient}
\end{figure}

The isotropic behaviour of the spin-pairs to magnetic fields facilitates a simplified construction of the gradient field device.
hBN nanopowder suspended in isopropanal is dropcast onto an RF waveguide embedded in the PCB with a magnet taped to one edge [Fig.~\ref{si_gradient}(a)].
The magnetic field strength decays moving away laterally from the magnet creating a field gradient [Fig.~\ref{si_gradient}(b)].
We collect ODMR images across the length of the waveguide to assign a spatial location to a given resonant frequency, providing a calibration for the device. 
We find the magnetic field decays from $\approx 0.2$\,T to $ \approx 0.05$\,T over $\approx 4$\,mm [Fig.~\ref{si_gradient}(c)], in good agreement with a numeric model of the magnetic field gradient (using the python package Magpylib\,\cite{OrtnerSoftwareX2020})

\section{Simulated Wi-Fi signal}

\begin{figure}[b]
    \centering
    \includegraphics{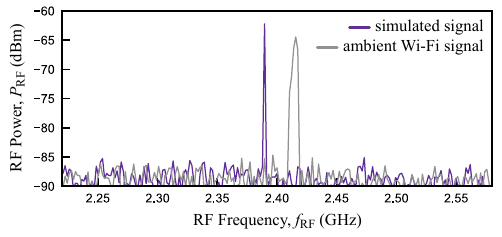}
    \caption{
    \textbf{Wi-Fi signal strength measurement.}
    Ambient (grey) and simulated (purple) Wi-Fi signals as seen by a traditional spectrum analyser connected to the receiving antenna. Note that the actual Wi-fi signal is intermittent.
    }
    \label{si_wifi}
\end{figure}

Here we compare the simulated Wi-Fi signal strength used for the test signal in Fig.~\ref{fig4}(b) to demonstrate analysis of real-world signals transmitted through free-space, to the signal strength of ambient Wi-Fi signal measured in the same room. 
A continuous signal at $2.38$\,GHz was output from the signal generator connected to a transmitting antenna and attenuated until the power collected by the receiving antenna matched the peak power of the observed Wi-Fi signal ($2.42$\,GHz) as seen on a traditional spectrum analyser [Fig.~\ref{si_wifi}].
The simulated signal was transmitted over $\approx 1$\,m.

\end{document}